\documentclass[aps,twocolumn,showpacs,amsmath,amssymb,prl,floatfix,superscriptaddress]{revtex4-1}
\usepackage{graphicx} % Include figure files
\usepackage{dcolumn}  % Align table columns on decimal point
\usepackage{color}
\usepackage{bm}
\usepackage{subfigure} % side-by-side figures
\graphicspath{{ps}}

\begin{document}

\title{\boldmath Precision measurement of charged pion and kaon differential cross sections in $e^+e^-$ annihilation at $\sqrt{s} = 10.52$~GeV}

\affiliation{University of the Basque Country UPV/EHU, 48080 Bilbao}
\affiliation{University of Bonn, 53115 Bonn}
\affiliation{Budker Institute of Nuclear Physics SB RAS and Novosibirsk State University, Novosibirsk 630090}
\affiliation{Faculty of Mathematics and Physics, Charles University, 121 16 Prague}
\affiliation{Justus-Liebig-Universit\"at Gie\ss{}en, 35392 Gie\ss{}en}
\affiliation{Gifu University, Gifu 501-1193}
\affiliation{Hanyang University, Seoul 133-791}
\affiliation{University of Hawaii, Honolulu, Hawaii 96822}
\affiliation{High Energy Accelerator Research Organization (KEK), Tsukuba 305-0801}
\affiliation{Ikerbasque, 48011 Bilbao}
\affiliation{University of Illinois at Urbana-Champaign, Urbana, Illinois 61801}
\affiliation{Indian Institute of Technology Guwahati, Assam 781039}
\affiliation{Indiana University, Bloomington, Indiana 47408}
\affiliation{Institute of High Energy Physics, Chinese Academy of Sciences, Beijing 100049}
\affiliation{Institute of High Energy Physics, Vienna 1050}
\affiliation{Institute for High Energy Physics, Protvino 142281}
\affiliation{Institute for Theoretical and Experimental Physics, Moscow 117218}
\affiliation{J. Stefan Institute, 1000 Ljubljana}
\affiliation{Kanagawa University, Yokohama 221-8686}
\affiliation{Institut f\"ur Experimentelle Kernphysik, Karlsruher Institut f\"ur Technologie, 76131 Karlsruhe}
\affiliation{Korea Institute of Science and Technology Information, Daejeon 305-806}
\affiliation{Korea University, Seoul 136-713}
\affiliation{Kyungpook National University, Daegu 702-701}
\affiliation{\'Ecole Polytechnique F\'ed\'erale de Lausanne (EPFL), Lausanne 1015}
\affiliation{Faculty of Mathematics and Physics, University of Ljubljana, 1000 Ljubljana}
\affiliation{Luther College, Decorah, Iowa 52101}
\affiliation{University of Maribor, 2000 Maribor}
\affiliation{Max-Planck-Institut f\"ur Physik, 80805 M\"unchen}
\affiliation{School of Physics, University of Melbourne, Victoria 3010}
\affiliation{Moscow Physical Engineering Institute, Moscow 115409}
\affiliation{Moscow Institute of Physics and Technology, Moscow Region 141700}
\affiliation{Graduate School of Science, Nagoya University, Nagoya 464-8602}
\affiliation{Kobayashi-Maskawa Institute, Nagoya University, Nagoya 464-8602}
\affiliation{Nara Women's University, Nara 630-8506}
\affiliation{National Central University, Chung-li 32054}
\affiliation{National United University, Miao Li 36003}
\affiliation{Department of Physics, National Taiwan University, Taipei 10617}
\affiliation{H. Niewodniczanski Institute of Nuclear Physics, Krakow 31-342}
\affiliation{Nippon Dental University, Niigata 951-8580}
\affiliation{Niigata University, Niigata 950-2181}
\affiliation{Osaka City University, Osaka 558-8585}
\affiliation{Pacific Northwest National Laboratory, Richland, Washington 99352}
\affiliation{Peking University, Beijing 100871}
\affiliation{Punjab Agricultural University, Ludhiana 141004}
\affiliation{Research Center for Electron Photon Science, Tohoku University, Sendai 980-8578}
\affiliation{RIKEN BNL Research Center, Upton, New York 11973}
\affiliation{Seoul National University, Seoul 151-742}
\affiliation{Sungkyunkwan University, Suwon 440-746}
\affiliation{School of Physics, University of Sydney, NSW 2006}
\affiliation{Tata Institute of Fundamental Research, Mumbai 400005}
\affiliation{Excellence Cluster Universe, Technische Universit\"at M\"unchen, 85748 Garching}
\affiliation{Toho University, Funabashi 274-8510}
\affiliation{Tohoku Gakuin University, Tagajo 985-8537}
\affiliation{Tohoku University, Sendai 980-8578}
\affiliation{Department of Physics, University of Tokyo, Tokyo 113-0033}
\affiliation{Tokyo Institute of Technology, Tokyo 152-8550}
\affiliation{Tokyo Metropolitan University, Tokyo 192-0397}
\affiliation{Tokyo University of Agriculture and Technology, Tokyo 184-8588}
\affiliation{CNP, Virginia Polytechnic Institute and State University, Blacksburg, Virginia 24061}
\affiliation{Wayne State University, Detroit, Michigan 48202}
\affiliation{Yamagata University, Yamagata 990-8560}
\affiliation{Yonsei University, Seoul 120-749}
  \author{M.~Leitgab}\affiliation{University of Illinois at Urbana-Champaign, Urbana, Illinois 61801}\affiliation{RIKEN BNL Research Center, Upton, New York 11973} % UIUC
  \author{R.~Seidl}\affiliation{RIKEN BNL Research Center, Upton, New York 11973} % RIKEN
  \author{M.~Grosse~Perdekamp}\affiliation{University of Illinois at Urbana-Champaign, Urbana, Illinois 61801}\affiliation{RIKEN BNL Research Center, Upton, New York 11973} % UIUC
  \author{A.~Vossen}\affiliation{Indiana University, Bloomington, Indiana 47408} % Indiana
  \author{I.~Adachi}\affiliation{High Energy Accelerator Research Organization (KEK), Tsukuba 305-0801} % KEK
  \author{H.~Aihara}\affiliation{Department of Physics, University of Tokyo, Tokyo 113-0033} % Tokyo
  \author{D.~M.~Asner}\affiliation{Pacific Northwest National Laboratory, Richland, Washington 99352} % PNNL
  \author{V.~Aulchenko}\affiliation{Budker Institute of Nuclear Physics SB RAS and Novosibirsk State University, Novosibirsk 630090} % BINP
  \author{T.~Aushev}\affiliation{Institute for Theoretical and Experimental Physics, Moscow 117218} % ITEP
  \author{A.~M.~Bakich}\affiliation{School of Physics, University of Sydney, NSW 2006} % Sydney
% ...
% BINP
\collaboration{The Belle Collaboration}
\noaffiliation

\begin{abstract}
Measurements of inclusive differential cross sections for charged pion and kaon production in $e^{+}e^{-}$ annihilation have been carried out at a center-of-mass energy of $\sqrt{s}=10.52$~GeV. The measurements were performed with the Belle detector at the KEKB $e^{+}e^{-}$ collider using a data sample containing $113 \times 10^6$ $e^+e^-\rightarrow q\overline{q}$ events, where $q=\{u,d,s,c\}$. We present charge-integrated differential cross sections $d\sigma_{h^{\pm}}/dz$ for $h^{\pm} = \{ \pi^{\pm},K^{\pm}\}$ as a function of the relative hadron energy $z=2E_h/\sqrt{s}$ from $0.2$ to $0.98$. The combined statistical and systematic uncertainties for $\pi^{\pm}$ ($K^{\pm}$) are $4\%$ ($4\%$) at $z \sim 0.6$ and $15\%$ ($24\%$) at $z \sim 0.9$. The cross sections are the first measurements of the $z$-dependence of pion and kaon production for $z>0.7$ as well as the first precision cross section measurements at a center-of-mass energy far below the $Z^0$ resonance used by the experiments at LEP and SLC. 
\end{abstract}


\maketitle

The fragmentation function (FF) $D_{q,\overline{q},g}^h(z,Q^2)$ parametrizes the transition of quarks $q$, antiquarks $\overline{q}$ and gluons $g$ into the color-neutral hadron $h$ in high-energy particle collisions as a function of the four-momentum transfer squared, $Q^2$, in the collision and the relative energy carried by the final state hadron, $z=2 E_h/\sqrt{s}$. FFs are assumed to be process-independent and are extracted through global analysis of inclusive hadron production in $e^{+}e^{-}$ and $pp$ reactions, and from semi-inclusive hadron production in $lN$ reactions at various center-of-mass energies (see Refs.~\cite{marcopaper,kumanopaper,akk}). Dokshitzer-Gribov-Lipatov-Altarelli-Parisi (DGLAP) evolution equations~\cite{dglap} derived from Quantum Chromodynamics (QCD) are used to jointly analyze data sets taken at different $Q^2$. The $Q^2$ evolution mixes quark and gluon degrees of freedom and enables the extraction of quark and gluon FFs through the analysis of precise hadron cross section data sets. In $e^{+}e^{-}$ annihilation, the availability of precision data sets was previously limited mostly to results from experiments (e.g., Refs.~\cite{aleph,delphi,opal,sld}) at LEP and SLC at the energy scale of the $Z^0$-boson mass. Recently, also the BaBar Collaboration reported inclusive hadron production cross sections at a center-of-mass energy of $10.54$~GeV~\cite{babarpaper}, however using a significantly smaller dataset compared to the Belle measurement.
The large data samples available at Belle complement existing results on hadron production cross sections with precise measurements at $Q^2=(10.52~\textnormal{GeV})^2$. The differential cross sections reported here are expected to improve the precision of FFs and the QCD-based determination of the nucleon quark- and gluon-structure from semi-inclusive deeply inelastic measurements at CERN, DESY and Jefferson Laboratory and inclusive proton-proton measurements at CERN and BNL. In addition, the Belle data will allow first quantitative tests of higher-order QCD effects at $z \sim 1.0$~\cite{werner}.

This analysis is based on a $68.0$~fb$^{-1}$ data sample collected with the Belle detector at the KEKB asymmetric-energy $e^+e^-$ ($3.5$ on $8$~GeV) collider~\cite{kekb} operating $60$~MeV below the $\Upsilon(4S)$ resonance at $\sqrt{s}=10.52$~GeV. The Belle detector is a large-solid-angle magnetic spectrometer that consists of a silicon vertex detector (SVD), a $50$-layer central drift chamber (CDC) and an array of aerogel threshold Cherenkov counters (ACC). The ACC detector is surrounded by a barrel-like arrangement of time-of-flight scintillation counters (ToF) and an electromagnetic calorimeter (ECL) comprised of CsI(Tl) crystals, located inside a superconducting solenoid coil that provides a $1.5$~T magnetic field. The iron flux-return yoke surrounding the coil is instrumented to detect $K_L^0$ mesons and to identify muons (KLM). The detector is described in detail elsewhere~\cite{belledetector}. The Belle Monte Carlo (MC) simulations are performed with the PYTHIA $6.205$~\cite{jetset} event generator and a GEANT$3$~\cite{geant} detector simulation.

An event must have at least three charged tracks and a heavy-jet mass (the invariant mass of the summed four-momenta of all tracks and ECL clusters in the jet with the largest invariant mass) at least above $1.8$~GeV/$c^{2}$ or above $25\%$ of the reconstructed visible energy. A reconstructed visible energy of more than $7$~GeV is required to reduce contamination from $\tau^{+}\tau^{-}$ events. From these events, we retain tracks with a laboratory-frame momentum $p_{\textnormal{lab}}$ of more than $500$~MeV/$c$, a scattering angle $\theta_{\textnormal{lab}}$ (relative to the beam axis) within the central detector acceptance ($-0.511 \leq \cos\theta_{\textnormal{lab}} < 0.842$), at least three SVD hits, and a point of closest approach to the $e^{+}e^{-}$ interaction point of under $4$~cm along the beam axis and under $1.3$~cm in the perpendicular plane. 

Using likelihood ratios based on $dE/dx$ measurements of the CDC, signals from the ACC, ToF measurements, ECL cluster energies as well as KLM signals, charged particles are identified as pions, kaons, protons, electrons or muons. The initially-measured track yield $N_{j,\textnormal{meas}}$ for particle type $j$ is binned in $z$ with width $0.01$, starting from $0.2$; negatively- and positively-charged tracks are treated separately. In order to obtain final cross sections, several corrections are applied, as described below. Statistical uncertainties are propagated through all corrections and systematic uncertainties from all corrections are added in quadrature. 

\begin{figure}[h]
	\begin{center}
		\includegraphics[width=0.45\textwidth]{{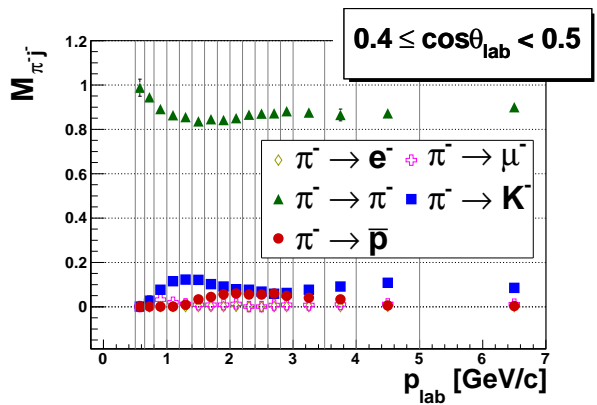}} %sampleprobs_afterextrapMConlyatloose_2_0_4_prl
		\caption{\label{fig:pid} (color online). Samples of extracted pion PID efficiencies (green triangles) and pion misidentification probabilities to kaons (blue squares), protons (red circles), muons (purple crosses) and electrons (green diamonds) for pions with $0.4 \leq \cos\theta_{\textnormal{lab}}<0.5$ as a function of the pion laboratory-frame momentum $p_{\textnormal{lab}}$. Electron probabilities are small and overlap with muon probability symbols in the plot.}  
	\end{center}
\end{figure}

The true yield $N_{i,\textnormal{true}}$ for particle type $i$ is obtained from the measured yields via
 \begin{equation}
N_{i,\textnormal{true}} = \sum_{j} M^{-1}_{ij} N_{j,\textnormal{meas}},
\label{pidcorr}
\end{equation}
where $M^{-1}$ is the inverse of the $5 \times 5$ particle identification (PID) probability matrix $M$, whose diagonal and off-diagonal elements represent PID efficiencies and misidentification probabilities, respectively. As the elements of $M$ depend on the laboratory-frame momentum and polar angle, an $M$ matrix is constructed in each of the $17\times9$ bins with $p_{\textnormal{lab}}$ boundaries $[0.5,0.65,0.8,1.0,\ldots,3.0,3.5,4.0,5.0,8.0)$~GeV/$c$ and $\cos\theta_{\textnormal{lab}}$ boundaries $[-0.511,-0.300,-0.152,0.017,$ $0.209,0.355,0.435,0.542,0.692,0.842)$. These values were chosen to obtain similar track yields in all bins. The elements of $M$, called ``efficiencies`` for the remainder of this letter, are obtained mostly from experimental data. For pions and kaons, the decay $D^{*+} \rightarrow \pi^+_{\textnormal{slow}} + D^0 \rightarrow \pi^+_{\textnormal{slow}} + (K^-\pi^+_{\textnormal{fast}})$ and its charge-conjugate are used. Similarly, proton PID efficiencies are extracted from $\Lambda \rightarrow p^+ + \pi^-$ and charge-conjugate decays and electron and muon efficiencies from $J/\psi \rightarrow e^+e^-,~\mu^+\mu^-$ decays. 
The procedures for calculating PID efficiencies from decays are rather similar and are described here for the pions in the $D^*$ channel. The reconstructed invariant mass distribution $m_{\pi_{\textnormal{slow}}(\pi_{\textnormal{fast}} K)} - m_{(\pi_{\textnormal{fast}} K)}$ for candidate triplets of tracks is fitted with a threshold background function and combinations of asymmetric Gaussian and polynomial functions for the signal. The initial yield of true pions in the sample can be obtained from the fitted yield of $D^{*}$'s in the invariant mass distribution before applying any PID selection criteria. Performing the fit on the same track sample but with additional PID selection criteria for pions, kaons, protons and leptons, and dividing by the initial yield of true pions, gives pion PID efficiencies. All fit uncertainties are propagated through the efficiencies, accounting for correlations. Several matrix elements, especially the ones for leptons due to the high $J/\psi$ mass relative to the quark energy, cannot be filled only from data-driven decay studies. Missing efficiencies are estimated once by extrapolating the trend of extracted probabilities with MC information and a second time by utilizing MC probabilities only, which produces two alternative sets of PID matrices; both are used. Examples of pion efficiencies are shown in Fig.~\ref{fig:pid}. Generally, the diagonal matrix elements have values around $90\%$ with the exception of kaon and proton efficiencies that drop at the highest momenta to about $70\%$ and $50\%$, respectively. The most prominent misidentification probabilities are $\pi \rightarrow K$ (up to $15\%$), $K\rightarrow \pi$ (up to $20\%$) and $p \rightarrow K$ (up to $50\%$).
For the correction according to Eq.~\ref{pidcorr}, both sets of PID matrices are inverted using singular value decomposition tools~\cite{svdbook}. Uncertainties on the elements of $M^{-1}$ are obtained by statistical analysis of large samples of inverted PID matrices whose elements are varied before inversion according to their uncertainties. Finally, these uncertainties are propagated as systematic uncertainties to the PID-corrected track yields. Measured yields $N_{j,\textnormal{meas}}$ are corrected with each set of alternative matrices. The final PID-corrected yields are calculated from the average of the two corrections. The difference between the corrections is assigned as additional systematic uncertainty. The PID correction changes pion (kaon) yields by around $+10\%$ ($-20\%$) at $z \sim 0.2$ and by around $+5\%$ ($+10\%$) at $z \sim 1.0$. The $z$-dependence of the changes is due to the kinematic dependence of the extracted PID efficiencies.

After the basic hadronic event selection, experimental data still contain hadron contributions from events other than $e^+e^-\rightarrow q\overline{q}$. A simulated sample containing all non-$q\bar{q}$ physically possible event topologies in $e^+e^-$ annihilation at $Q^2=(10.52~\textnormal{GeV})^2$ is assembled from dedicated QED (BHLUMI~\cite{bhlumi} generator for Bhabha-scattering, KKMC~\cite{kk} generator for $\mu$ and $\tau$ pair production) and two-photon MC data. Hadron yields from such non-$q\bar{q}$ events passing the hadronic event selection are scaled to the luminosity of the measurement sample and subtracted from the PID-corrected yields. Contributions from events other than $e^+e^-\rightarrow q \overline{q}$ are found not to be significant at $z \sim 0.2$. At $z \sim 1.0$, about $30\%$ of the reconstructed pions and $10\%$ of the reconstructed kaons are created in non-QCD $\tau\tau$ and two-photon events. Statistical uncertainties are propagated as systematic uncertainties on the corrected yields.

The reconstructed values of $z$ are smeared around the true values because of the non-zero momentum resolution, which is smaller than half of the constant bin width of $0.01$ at $z \sim 0.2$ and comparable to the bin width at $z \sim 0.9$. Similar to the particle misidentification correction, the $z$-bin migration can be described by a matrix. While a $z$ cut-off of $0.2$ is applied to the final cross sections, the smearing matrix needs to be evaluated for $z<0.2$ to take into account smearing into or out of the selection region. Smearing matrices are extracted down to $z~=~0.15$ for pions and $z~=~0.17$ for kaons from the MC simulation. The matrices are inverted using a regularized unfolding procedure described in Ref.~\cite{regularization}. Due to limitations of the method, the correction is only applied to yields below $z~=~0.98$ for pions and kaons and so imposes an upper limit on $z$ for the final cross sections. All uncertainties are propagated through the unfolding procedure. Additional systematic uncertainties are assigned from tracks smearing out of and into the interval $0.2 \leq z < 0.98$, the choice of the regularization weights and the difference in momentum resolution of the detector simulation compared to the resolution extracted from experimental cosmic ray data given in Ref.~\cite{belledetector}. The smearing correction causes no significant change in the measured yields because of the good momentum resolution, small bin width and mostly symmetric smearing to lower and higher values of $z$. 

Some hadrons are lost or gained through decays in flight. In addition, hadrons are lost due to interaction with detector material while others are created in such interactions. Finally, the track finding and fitting algorithm may find fake tracks or fail to reconstruct true tracks because of detector inefficiencies or occupancy effects. Limitations in the MC simulation do not allow us to distinguish these processes. Thus, a joint correction is performed for all processes in each $z$ bin, extracting ratios between reconstructed and generated particle yields. 
Concerning decay-in-flight (DIF), the QCD framework used by global analyses usually does not account for hadrons created in weak decays, which are nevertheless present in measured cross sections. Correction factors to remove pions and kaons produced in weak decays (e.g., from $\tau$ leptons, kaons and $D$ mesons, $\Lambda$ and heavier baryons) would be purely simulation-dependent. To be consistent, all weakly produced pions and kaons are included in the results, and all decayed pions and kaons are recovered. For the final cross sections, we provide supplementary fractions of pions and kaons originating from strong and weak decays~\cite{archiveddata}, extracted from MC data. For $z \lesssim 0.5$, the fraction of weakly-produced pions (kaons) is $30\%$ ($50\%$) but vanishes towards $z \sim 1.0$ due to phase-space limitations.
The implementation of DIF in the MC simulation is tested by analytical calculations and found to be consistent. Correlation effects between the PID and DIF corrections are studied; the observed correlations are applied as additional systematic uncertainties. The effect of hadronic interaction modeling on particle yields is estimated by comparing efficiencies from the default FLUKA~\cite{fluka} with the GEISHA~\cite{geisha} package and with no hadronic interactions. The small remaining differences are assigned as systematic uncertainties. Fake- and multiple-track reconstruction is found to be negligible. Track reconstruction inefficiencies are at the few-percent level and are corrected for. The most significant track quality cut is the requirement for three or more hits in the SVD, which also introduces most of the dependence of the efficiencies in this correction for $z \gtrsim 0.7$. The overall efficiencies remain above $85\%$ for all $z$. Statistical and systematic uncertainties on the efficiencies are propagated as systematic uncertainties on the corrected yields.

To obtain the best-measured reconstructed tracks, the analysis is limited to the barrel region of the detector. A correction is applied to recover the $4\pi$ physical particle yield. Corresponding efficiencies from the MC simulation decrease from about $65\%$ at $z \lesssim 0.5$ and level out around $60\%$ at $z \sim 1.0$. The behavior at $z \sim 1.0$ is consistent with the assumption that hadrons with $z \sim 1.0$ follow the known $1 + \cos^2 \theta_{\textnormal{cms}}$ distribution of fragmenting quarks, where $\theta_{\textnormal{cms}}$ is the scattering angle in the center-of-mass frame. For $z < 1.0$, additional transverse hadron momentum causes the $\cos\theta_{\textnormal{cms}}$ spectra to become constant and efficiencies to increase. The MC description of the $\cos\theta_{\textnormal{cms}}$ dependence in the fragmentation process is tested with experimental data yields in a two-dimensional $z$ and $\cos\theta_{\textnormal{cms}}$ binning. Consistency is found within statistical uncertainties and no additional systematic uncertainty is assigned. Statistical uncertainties on the efficiencies are propagated as systematic uncertainties on the corrected yields.

Hard initial-state and final-state photon radiation (ISR/FSR) processes reduce the fragmentation energy scale $\sqrt{s}/2$ for both (ISR) or for one (FSR) final-state quark. Therefore, experimentally measured yields contain a variety of fragmentation scales $\sqrt{s}/2$. A theoretical deconvolution of ISR/FSR from measured yields is beyond the scope of this measurement. Instead, we keep the energy scales in the measurement sample within $0.5\%$ of the nominal $\sqrt{s}/2$ in order to remain below the scale resolution of the state-of-the-art next-to-leading order DGLAP evolutions. Corresponding fractions of hadrons from events with summed ISR/FSR photon energies of less than $0.5\% \times \sqrt{s}/2$ are extracted from the MC simulation for each $z$ bin. These fractions rise from $65\%$ at $z \lesssim 0.5$ to almost $100\%$ at maximum $z$ due to phase-space limitations for ISR/FSR. The fractions are applied bin-by-bin to the measured yields to exclude particles from events with large ISR/FSR contributions. According to MC simulations, $35\%$ of all events are excluded. Systematic uncertainties are assigned from the dependence of the hadron fractions on the chosen PYTHIA MC parameter sets and from accounting for the remaining scale variance in the sample. 

Efficiencies for all applied event selections are extracted from MC simulations. Efficiencies for the hadronic event selection are close to unity at $z \sim 0.2$. They drop rapidly at larger $z$ since both the track multiplicity and heavy jet mass requirements disfavor events containing tracks with $z \sim 1.0$. In such events, the phase space for additional particles and substantial remaining transverse momentum is significantly reduced. The efficiencies drop to about $60\%$ at $z \sim 0.9$ and to less than $10\%$ for $z \sim 1.0$. Conversely, efficiencies for the visible energy requirement are above $95\%$ at $z \sim 1.0$, but drop to about $75\%$ for particles with $z \sim 0.2$. This can be understood from the increased likelihood for events containing tracks with $z \sim 1.0$ to deposit sufficient energy in the barrel part of the detector. The measured yields are corrected accordingly and statistical uncertainties are propagated as systematic uncertainties on the corrected yields. The distributions of event shape variables used in the event selection are compared between MC and experimental data. Differences in these distributions causing variations in event selection efficiencies are assigned as systematic uncertainties.

\begin{figure}[h]
	\begin{center}
		\includegraphics[width=0.45\textwidth]{{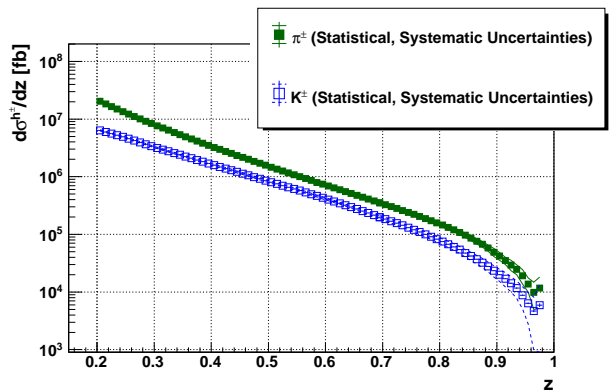}} %correctiondiagnostics_crosssectnormalizedj062313I_chargeinteg
		\caption{\label{fig:final} (color online). Final charge-integrated differential pion (green solid boxes) and kaon (blue empty boxes) production cross sections as a function of the fractional hadron energy $z$. Statistical uncertainties are shown by error bars, systematic uncertainties by error bands.}  
	\end{center}
\end{figure}

After all corrections, the measured yields are normalized to the time-integrated luminosity of the measurement sample, $68.0$~fb$^{-1}$, with an uncertainty of $1.4\%$. The resulting final charge-integrated differential pion and kaon cross sections $d\sigma_{h^{\pm}}/dz$ for $h^{\pm} = \{ \pi^{\pm},K^{\pm}\}$ are displayed in Fig.~\ref{fig:final}. All cross section values and uncertainties are available in Ref.~\cite{archiveddata}. Combined relative statistical and systematic uncertainties remain below $5\%$ up to fractional hadron energies $z \sim 0.65$, then rise to about $15\%$ (pions) and $24\%$ (kaons) at $z \sim 0.9$ and reach $55\%$ (pions) and more than $100\%$ (kaons) at maximum $z$. As a test of all applied corrections, pion and kaon charge ratios $R_{i} = N_{i^{-}}/N_{i^{+}}$ are fitted with a constant. The results, $R_\pi = 0.995 \pm 0.008$ and $R_K = 1.000 \pm 0.009$ (with a combined statistical and systematic uncertainty), are consistent with $1.0$ within the extracted uncertainties and indicate consistency of the performed corrections.
The precision of the present measurement is systematics-limited for all $z$. For $z \lesssim 0.5$, the dominant systematic uncertainties arise from the initial/final state radiation correction. At $z \sim 1.0$, the momentum smearing, particle identification and DIF/reconstruction corrections represent the largest contributions to the systematic uncertainties. Figures showing different contributions to the systematic uncertainties for the final pion and kaon cross sections are given in Ref.~\cite{archiveddata}.

\begin{figure}[h]
	\begin{center}
			\includegraphics[width=0.45\textwidth]{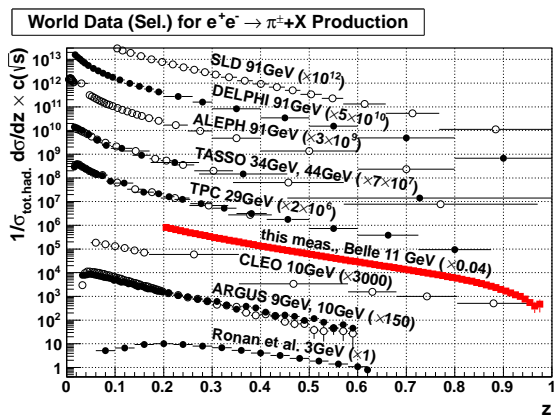} %c1x1_pi_normj062313I
		\caption{\label{fig:worlddata} (color online). Belle results compared to charge-integrated normalized pion cross sections (multiplicities) from selected measurements. For better visibility, all data sets are scaled by arbitrary numbers (indicated between parentheses)~\footnote{Presentation style based on plots created by O.~Biebel {\it et al.} shown in Ref.~\cite{ffreview}}. Statistical and systematic uncertainties are added in quadrature.}  
	\end{center}
\end{figure}

The final cross sections are compared with normalized cross section measurements from LEP and SLC experiments as well as from other, lower energy $e^{+}e^{-}$ experiments. As an illustration, the pion cross section is compared to those from Refs.~\cite{aleph,delphi,sld,argus9and10,cleo10,slac3,tasso34and44,tpc29prl52,tpc29prl61} in Fig.~\ref{fig:worlddata}. The corresponding plot for kaon cross sections can be found in Ref.~\cite{archiveddata}. The resolution in $z$ is significantly improved for all compared normalized cross sections over most of the $z$ range of this measurement. In addition, no other previous measurement probes the $z$ dependence of hadron production for $z \gtrsim 0.7$. The total relative uncertainties of the previous measurements described above are larger or comparable to the uncertainties achieved in the results reported here. In particular, significantly better precision than previous measurements at low energy scales is reached. A comparison of the Belle results with simulated data shows agreement for $z \lesssim 0.5$, but exhibits a strong dependence on the chosen simulation parameter values at $z \gtrsim 0.6$ (see Ref.~\cite{archiveddata}).

In conclusion, differential cross sections of identified charged pion and kaon production are measured over a broad range in $z$ with $0.2 \leq z < 0.98$ with high relative precision. The analyzed data sample of $68.0$~fb$^{-1}$ has been accumulated at a center-of-mass energy of $\sqrt{s} = 10.52$~GeV, rendering this measurement the first precision measurement far from LEP/SLC center-of-mass energies. The high statistics and good control of systematic uncertainties will, for the first time, give constraints on the dependence of hadron FFs at $z \gtrsim 0.7$ and allow studies of higher order QCD effects at $z \sim 1.0$.
\begin{acknowledgments}
The authors would like to thank M. Stratmann, D. de Florian, R. Sassot, S. Kumano, F. Jegerlehner, S. Jadach and H. Czyz for valuable discussions and suggestions. We thank the KEKB group for excellent operation of the accelerator; the KEK cryogenics group for efficient solenoid operations; and the KEK computer group, the NII, and PNNL/EMSL for valuable computing and SINET4 network support. We acknowledge support from MEXT, JSPS and Nagoya's TLPRC (Japan); ARC and DIISR (Australia); NSFC (China); MSMT (Czechia); DST (India); INFN (Italy); MEST, NRF, GSDC of KISTI, and WCU (Korea); MNiSW and NCN (Poland); MES and RFAAE (Russia); ARRS (Slovenia); IKERBASQUE and UPV/EHU (Spain); SNSF (Switzerland); NSC and MOE (Taiwan); and DOE and NSF (USA).
\end{acknowledgments}
\end{document}